\documentclass[conference]{IEEEtran}
\usepackage{booktabs} 
\usepackage{multirow}
\usepackage{cite}
\usepackage{amsmath,amssymb,amsfonts}
\usepackage{algorithmic}
\usepackage{graphicx}
\usepackage{textcomp}
\usepackage{xcolor}
\usepackage{cleveref}
\usepackage{siunitx}
\usepackage{microtype}
\usepackage{subcaption}

\begin{document}

\newcommand\atlas{{\textsc{Atlas}}}
\newcommand\tnmt{{\textsc{TestNMT}}}
\newcommand\reassert{{\textsc{ReAssert}}}
\newcommand\reformer{{Reformer}}
\newcommand\tct{{\textsc{TCTracer}}}
\newcommand\tabspace{\vspace{-0.5ex}}
\newcommand\figbotspace{\vspace{-1ex}}

\newcommand{\ROT}[1]{\rotatebox[origin=c]{90}{#1}}

\title{\reassert{}: Deep Learning for Assert Generation}

\author{
\IEEEauthorblockN{Robert White\IEEEauthorrefmark{1}, Jens Krinke\IEEEauthorrefmark{2}}
\IEEEauthorblockA{\IEEEauthorrefmark{1}University College London \emph{robert.white.13@ucl.ac.uk}}
\IEEEauthorblockA{\IEEEauthorrefmark{2}University College London \emph{j.krinke@ucl.ac.uk}}
}

\author{\IEEEauthorblockN{Robert White}
\IEEEauthorblockA{\textit{University College London}\\
robert.white.13@ucl.ac.uk}
\and
\IEEEauthorblockN{Jens Krinke}
\IEEEauthorblockA{\textit{University College London}\\
j.krinke@ucl.ac.uk}
}

\maketitle

\begin{abstract}
The automated generation of test code can reduce the time and effort required to build software while increasing its correctness and robustness.
In this paper, we present \reassert{}, an approach for the automated generation of JUnit test asserts which produces more accurate asserts than previous work with fewer constraints.
This is achieved by targeting projects individually, using precise code-to-test traceability for learning and by generating assert statements from the method-under-test directly without the need to write an assert-less test first.
We also utilise \reformer{}, a state-of-the-art deep learning model, along with two models from previous work to evaluate \reassert{} and an existing approach, known as \atlas{}, using lexical accuracy, uniqueness, and dynamic analysis. 
Our evaluation of \reassert{} shows up to 44\% of generated asserts for a single project match exactly with the ground truth, increasing to 51\% for generated asserts that compile.
We also improve on the \atlas{} results through our use of \reformer{} with 28\% of generated asserts matching exactly with the ground truth.
\reformer{} also produces the greatest proportion of unique asserts (71\%), giving further evidence that \reformer{} produces the most useful asserts.
\end{abstract}

\begin{IEEEkeywords}
  software engineering, software testing, test generation, machine learning
\end{IEEEkeywords}

\section{Introduction}
\label{intro}
The process of creating and maintaining unit tests is time-consuming, error-prone, and often disliked by developers, frequently resulting in software that has a low level of test coverage.
Previous work has shown that to maintain a high level of unit test coverage, the tests must be created at the same time as the tested code as retroactively creating unit tests is rarely done and only partially successful when attempted~\cite{klammer2015writing}.
Therefore, by automating parts of the unit test creation process we hope to improve the efficiency of the software engineering process and the robustness of the resulting software.
To achieve this, we present \reassert{}, an approach for generating JUnit assert statements using deep learning.

Test suite generation tools such as EvoSuite~\cite{fraser2013whole}, Randoop~\cite{pacheco2007randoop}, and AgitarOne~\cite{agitar} employ techniques that primarily focus on generating high-coverage tests rather than meaningful asserts and, therefore, the asserts they generate are often weak and lack specificity.
This problem contributes to the deficiencies these tools show when attempting to revealing real-world faults~\cite{shamshiri2015automated, shamshiri2015do}.

To overcome the issues that existing test generation techniques have with generating asserts, we turn to deep learning.
Previous work~\cite{white2018testnmt} investigated generating JUnit tests using a sequence to sequence recurrent neural network (RNN) trained on individual projects in an approach named \tnmt{}.
After this, Watson et al.~\cite{watson2020on} used a similar RNN model in their \atlas{} approach, trained on a general corpus mined from GitHub for generating just the assert statements for JUnit tests.
These previous works have demonstrated that this type of deep neural network is capable of generating useful test code, however, in the case of \tnmt{}, the generated candidate tests need some manual transformation before being usable and, in the case in case of \atlas{}, only 17\% of the generated asserts were exact matches with the ground truth when using the raw dataset and the test (minus the asserts) was required to already been written.
In addition, only a single assert could be generated for a given method and no analysis beyond lexical accuracy was performed to assess the usefulness of the asserts.

Our approach, called \reassert{}, builds on the previous work by focusing on generating JUnit asserts, similar to Watson et al.~\cite{watson2020on}, but utilises a project-based approach that does not require the (assert-less) test to be written before asserts can be generated and allows for the generation of more than one assert per tested method.
\reassert{} can use three different models and includes the new \reformer{} model~\cite{kitaev2020reformer} in addition to the two RNN models used in \tnmt{}~\cite{white2018testnmt} and \atlas{}~\cite{watson2020on}.
\reformer{} utilises a state-of-the-art deep learning architecture and may push the accuracy and usefulness of the generated asserts beyond that of the previous models.
All three models are applied to both \reassert{} and a re-implementation of \atlas{}, and we also expand the evaluation in two other ways to focus more on real-world usefulness and applicability.
Firstly, we perform an extended lexical accuracy evaluation (how close is the text of the generated asserts to the ground truth from the test set) and an analysis of the uniqueness of the generated asserts, which gives further evidence as to their usefulness.
Secondly, for \reassert{}, we go beyond the static lexical accuracy analysis, to use a dynamic analysis which determines how many generated asserts compile and how many pass when inserted into existing tests.
By evaluating all three models for both \reassert{} and \atlas{} with the uniqueness and dynamic evaluation, along with the typical lexical evaluation, we demonstrate which approach and model combinations are the most useful in a real-world setting.
The main contributions of this paper are:
\begin{itemize}
  \item \reassert{}, a project-based deep learning approach for the generation of unit test asserts implemented for JUnit.
  \item An evaluation of \reassert{} using lexical accuracy and dynamic analysis with \reformer{}, a new state-of-the-art transformer-based model and two RNN-based models from previous work.
  \item An extended comparative evaluation of all three models using lexical accuracy and uniqueness on a previous approach, \atlas{}.
  \item Takeaway messages for researchers and practitioners concerning the construction of data sets when applying sequence to sequence learning for code generation.
\end{itemize}

\label{sec:approach}
\begin{figure*}
  \centering
  \includegraphics[width=\textwidth]{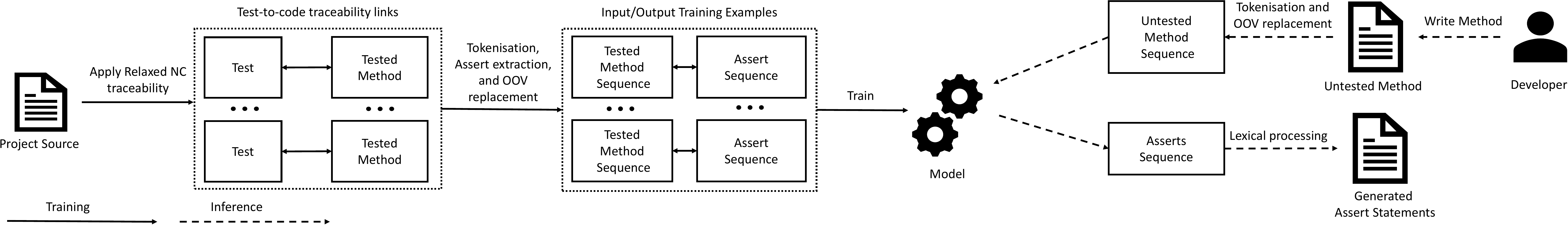}
  \vspace{-4ex}
  \caption{Overview of the \reassert{} approach.}\figbotspace
  \label{fig:reassert-approach}
\end{figure*}

\section{Background}
\label{sec:background}
Test assert generation has previously been the domain of test suite generation tools, such as EvoSuite~\cite{fraser2013whole}, Randoop~\cite{pacheco2007randoop}, and AgitarOne~\cite{agitar}.
However, these tools primarily generate tests through methods that optimise for coverage, such as genetic programming and random testing.
Therefore, these tools produce tests that aim primarily to achieve high coverage rather than include meaningful assert statements, resulting in a deficiency in the ability of these generated tests to detect real faults.
This was quantified in a study~\cite{shamshiri2015automated} which discovered that neither EvoSuite, Randoop, or AgitarOne were able to detect more than 40.6\% of faults in the Defects4J~\cite{just2014defects4J} database.
As 63.3\% of the undetected faults were covered, this indicates weaknesses in the asserts of the generated test cases.

With the advent of deep learning and the successful application of deep learning techniques to tasks that require the processing of sequential data, especially language-based tasks such as machine translation~\cite{kalchbrenner2013recurrent,sutskever2014sequence,cho2014on}, an opportunity to apply these methods to source code was created.
These deep learning models have been applied to a wide range of software engineering problems such as code summarisation~\cite{allamanis2016convolutional,alon2018code2seq,iyer2016summarizing}, program comprehension~\cite{henkel2018code}, clone detection~\cite{white2016deep}, code similarity~\cite{zhao2018deepSim}, method name generation~\cite{alon2019code2vec}, comment generation~\cite{hu2018deep}, traceability~\cite{guo2017semantically}, and type inference~\cite{hellendoorn2018deep}.
However, for the task of code generation, deep learning models initially were only applied to the generation of implementation code~\cite{ling2016latent}, not test code.
\tnmt~\cite{white2018testnmt} applied these techniques to test generation by utilising a sequence to sequence RNN-based neural network, adapted from a model that had previously been used for neural machine translation and applied it to translate from Java methods to JUnit tests.
\tnmt{} demonstrated that, when applied to large individual projects, this technique is capable of generating some tests that only require a small amount of manual effort on the part of the developers to turn into useful tests.
However, many tests still required a large amount of effort to be converted and the approach was not effective when using a single multi-project data set to train a general model that works for any project.
After \tnmt, \atlas~\cite{watson2020on} applied the same type of model to the problem of generating test code, however, instead of attempting to generate whole tests, \atlas{} attempts only to generate the asserts for JUnit test cases.
This removes the issue of developers having to expend a lot of effort to transform the output of the model into usable code and also allows the training of a single network on a corpus of general Java code to apply to any project.
However, unlike \tnmt~\cite{white2018testnmt}, \atlas{} only uses tests with a single assert statement and the test code (minus the assert) is included in the source sequence, requiring that a developer writes a test before using \atlas, which can then only generate a single assert.
The model used by \tnmt~\cite{white2018testnmt} and the model used by \atlas~\cite{watson2020on} are utilised in this work for a comparative evaluation with \reassert{} and the \reformer{} model.

\section{Approach}
\begin{figure}
  \centering
  \includegraphics[width=\columnwidth]{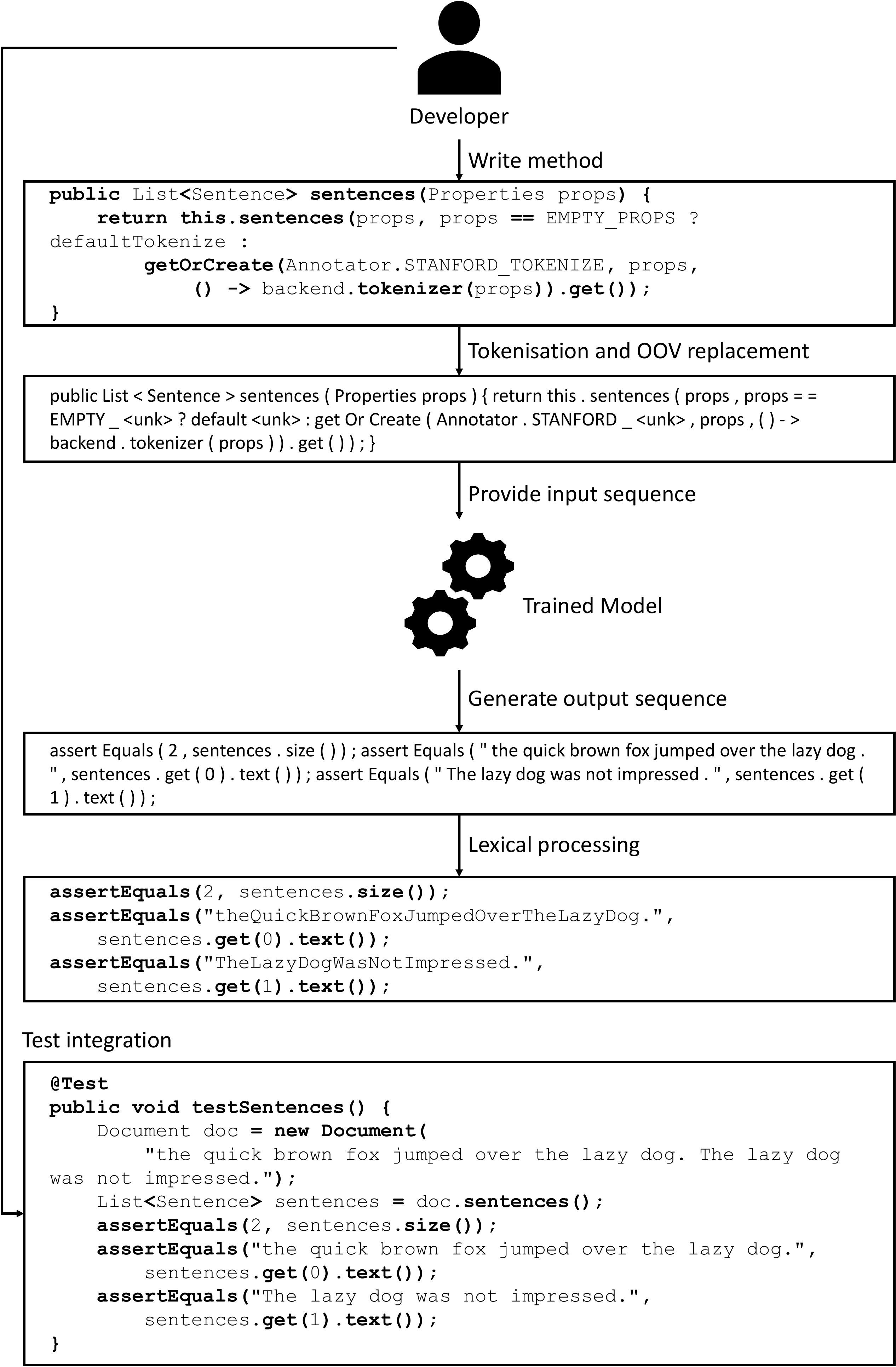}
  \vspace{-2ex}
  \caption{Example from the Stanford CoreNLP project demonstrating the \reassert{} process to generate asserts for a method.}
  \label{fig:reassert-approach-example}\figbotspace
\end{figure}

The \reassert{} approach, illustrated in \Cref{fig:reassert-approach}, facilitates the generation of assert statements for a given method by using deep neural network models trained on pairs of assert statements and tested methods, extracted from existing test-to-tested-method pairs.
To train the model, we start by gathering the test-to-tested-method pairs from a target project via test-to-code traceability links~\cite{rompaey2009establishing}.
Then, for each test-to-tested-method pair, we extract the assert statements from the test method and concatenate them to produce the string of assert statements associated with the tested method.
The tested method and assert strings are then processed into input and output token sequences, known as method sequences and assert sequences respectively.
These sequences are used to train the model.
Once trained, the model can be used to generate an assert sequence, given a method sequence as input.
The generated assert sequences are then processed into syntactically correct code that can be directly inserted into a test for that method.
\Cref{fig:reassert-approach-example} illustrates an example from Stanford CoreNLP for how \reassert{} generates asserts for a new method by processing the method into an input sequence, inferring over the trained model, and processing the output sequence into syntactically correct asserts.
As the example shows, the generated assert statements can easily be expanded into a test.

We specifically target our work more toward applicability than previous work by ensuring that we do not apply any filtering or abstraction and we do not use any prediction techniques that generate multiple outputs, such as beam search.
We also only use the training set to create the vocabulary that is used when generating the method sequences and assert sequences.
This is to ensure, firstly, that we are evaluating in a scenario that is true to real-world development and, secondly, that we minimise the amount of work that a developer has to do to utilise the produced assert statements in their code.
Also, in contrast to \atlas{}, \reassert{} does not include the test code in the source sequence as this would require the developer to have already written the test case (except for the assert statements) before using \reassert{} to generate asserts.
As we believe that the generated asserts should help the developer to write the rest of the test, this is an important improvement over prior work.
Our use of tests which have multiple assert statements is another improvement over \atlas{} which only uses tests that contain a single assert statement.
We believe this further increases \reassert's applicability.

\subsection{Test-to-code Traceability Establishment}
\label{subsec:approach-test-to-code-traceability-establishment}
Given the code for a project, we first need to extract the test-to-code traceability links in order to build our training and testing data sets.
Establishing test-to-code traceability links is an open research problem in software engineering for which multiple different techniques have been developed.
Each technique has its own strengths and weaknesses, resulting in different balances between precision and recall~\cite{rompaey2009establishing,white2020establishing}.
Finding the right balance of precision and recall is important for building a data set for machine learning as if the precision is too low, the data will have too much noise (incorrect links) but if the recall is too low the data set will be too small to effectively train from.
In addition, the optimal precision versus recall trade-off differs depending on which data set we are constructing.
When constructing the training set, we prefer recall, however, for the validation set (used for configuring the parameters of the networks) and test set (used for the evaluation) we prefer precision.
This is due to the fact that when we are training the model we want to ensure we have as much data as possible, whereas, when we are evaluating the model using the validation or test set, we want to ensure that we are not evaluating the model with noisy data.

When training a model, we can tolerate some noise in the data as we want to maximise the amount of data and, even if a link is technically incorrect, we may still be able to learn some useful structure from it.
An example of this can be seen when looking at tests for commonly overridden methods such as \textit{equals} or \textit{toString}.
In these cases, even if a link is incorrect, a link between the test for the \textit{equals} method of one class to the \textit{equals} method of a different but similar class, the network may still learn some useful information about the general structure of \textit{equals} tests because most tests for \textit{equals} methods tend to be very similar.
However, when we are evaluating the model, we want to ensure that there are as few incorrect links as possible as we don't want to be evaluating the model by asking it to generate something that is incorrect.
Doing so will give an inaccurate view of how well the model performs, usually resulting in an underestimation of its accuracy.

Given the above concerns, we firstly want to find a high precision technique for building the validation and training sets.
Using the recent work by White et al. which compares the precision and recall of multiple techniques~\cite{white2020establishing}, we selected the naming conventions (NC) technique for building these data sets as it has a precision of 100\%.
The naming convention technique establishes links by taking the fully-qualified names (FQNs) of both the tested method and the test method and comparing them after the word \textit{test} has been removed from the test method name.
If the names match exactly, the test method is linked to the tested method.
However, given the very low recall of only 11\%, this technique was not suitable for building the training set so we created a variant of this technique called ``Relaxed NC'' (we conversely call the standard NC ``Strict NC'').
Relaxed NC utilises the same concept as Strict NC but instead of performing the matching on the FQN, the matching is performed only over the tested method and test method name. Therefore, a link will still be created even if the class name does not match the test class name.
While this will create some incorrect links, we can often still learn some useful structure from these links (as described above) and it gives us much more data to train on, which is critical, especially for the smaller projects.

\subsection{Data Set Construction}
\label{subsec:data-set-construction}
To build the data sets, we start by constructing all the Strict NC and Relaxed NC links and place all the Relaxed NC links into the training set.
The Strict NC links are then split between the validation set (used for configuring the parameters of the networks) and test set (used for the evaluation), up to the maximum size of 100 links for each set. 
Any excess Strict NC links are placed in the training set.
As we want to ensure that we are not unfairly biasing the model, we then filter out any links from the training set that appear in either the validation or the test sets.
This filtering can result in a large reduction of the number of links in the training set, with the number of links in the validation and test set greatly influencing the magnitude of this reduction as the larger the validation and test sets are, the more links will have to be removed from the training set. 
Therefore, it is important to balance the sizes of the sets so that each set has an adequate amount of links to perform its function.
This is why we limit the number of links in the test and validation sets to 100. 
Now that we have the links for each set we must process the links into pairs of source sequences (from the tested methods) to target sequences (from the tests).
To do this we first tokenise the source for each artefact, build the vocabularies based on the tokenised sequences, then replace out-of-vocabulary (OOV) tokens with UNK.
Our tokenisation process consists of stripping all non-printing and non-alphanumeric characters that are not used in Java, adding spaces around all programming language characters, de-camelcasing identifiers and adding spaces around the resulting tokens.
This results in a sequence of individual tokens consisting of the split identifiers and programming language characters.
At this stage the tested methods have been fully tokenised into source sequences, however, for the tests we want to keep only the assert statements so we add another stage of processing for the tests where detect which tokens are part of assert statements by checking for the "assert" token, finding the next opening parenthesis and its partner closing parenthesis and treating all the tokens in-between as being part of the assert statement.
Any tokens that are not determined to be part of an assert statement are deleted.
This results in a target sequence that is just the tokenised assert statements for the test.
It is important to note that unlike Watson et al., we use tests that have multiple assert statements and we add all of the assert statements in the test to the target sequence.
Once we have applied this process to all the code snippets for the tested methods and the tests we have our sets of input-output examples (source sequence to target sequence pairs).
We then build the source and target vocabularies by collecting all the tokens in all the sequences and taking the top $n$ most frequent tokens, where $n$ is the desired size of the vocabulary.
The vocabularies are then used to replace OOV tokens in the sequences by replacing any token which does not appear in the relevant vocabulary with UNK.

\begin{figure*}
  \centering
  \begin{subfigure}{0.33\linewidth}
    \centering
    \includegraphics[width=0.85\columnwidth]{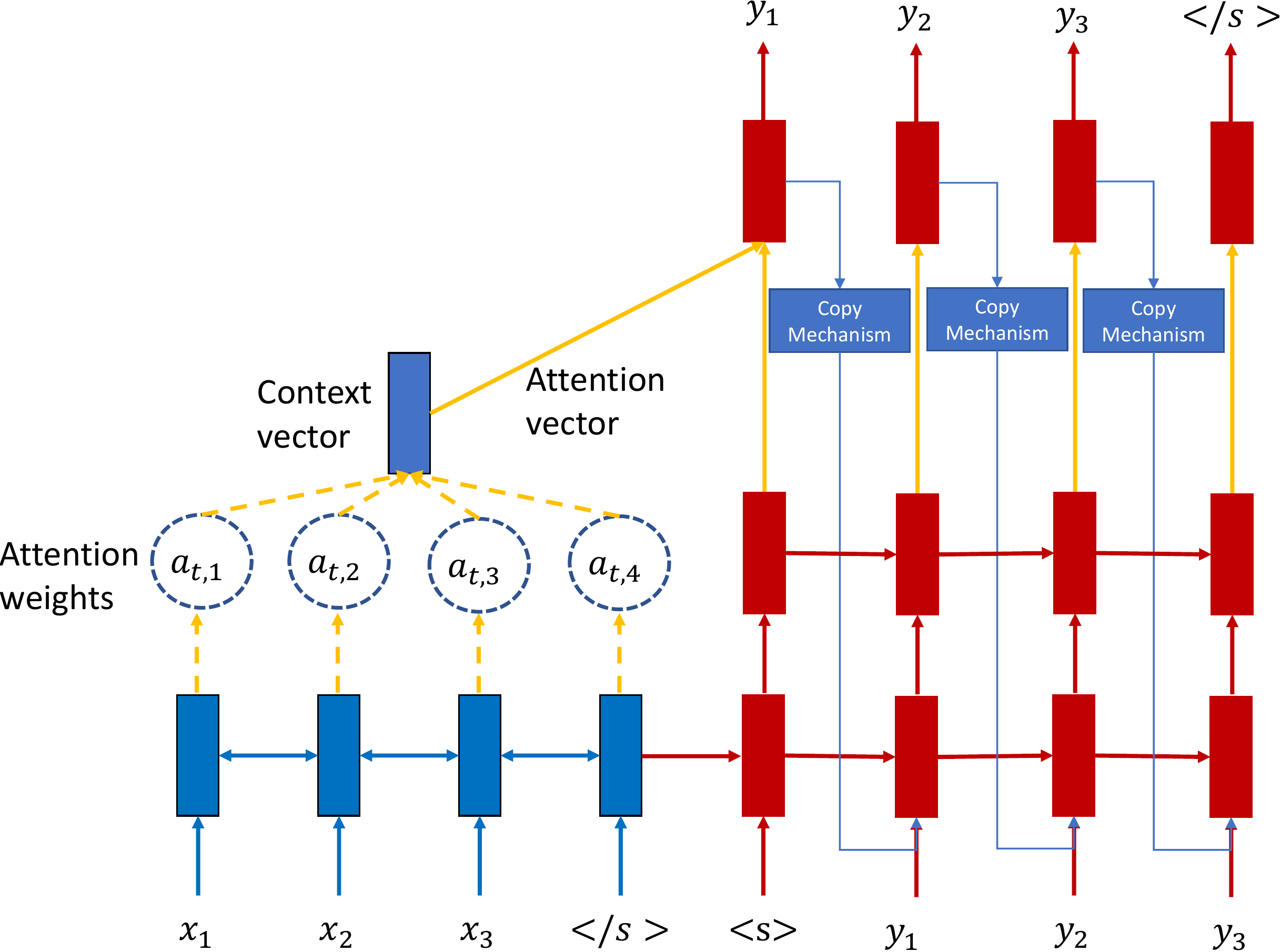}
    \caption{\atlas{}{\tiny/RNN}}
    \label{fig:atlas-architecture}
  \end{subfigure}%
  \begin{subfigure}{0.33\linewidth}
    \centering
    \includegraphics[width=0.85\columnwidth]{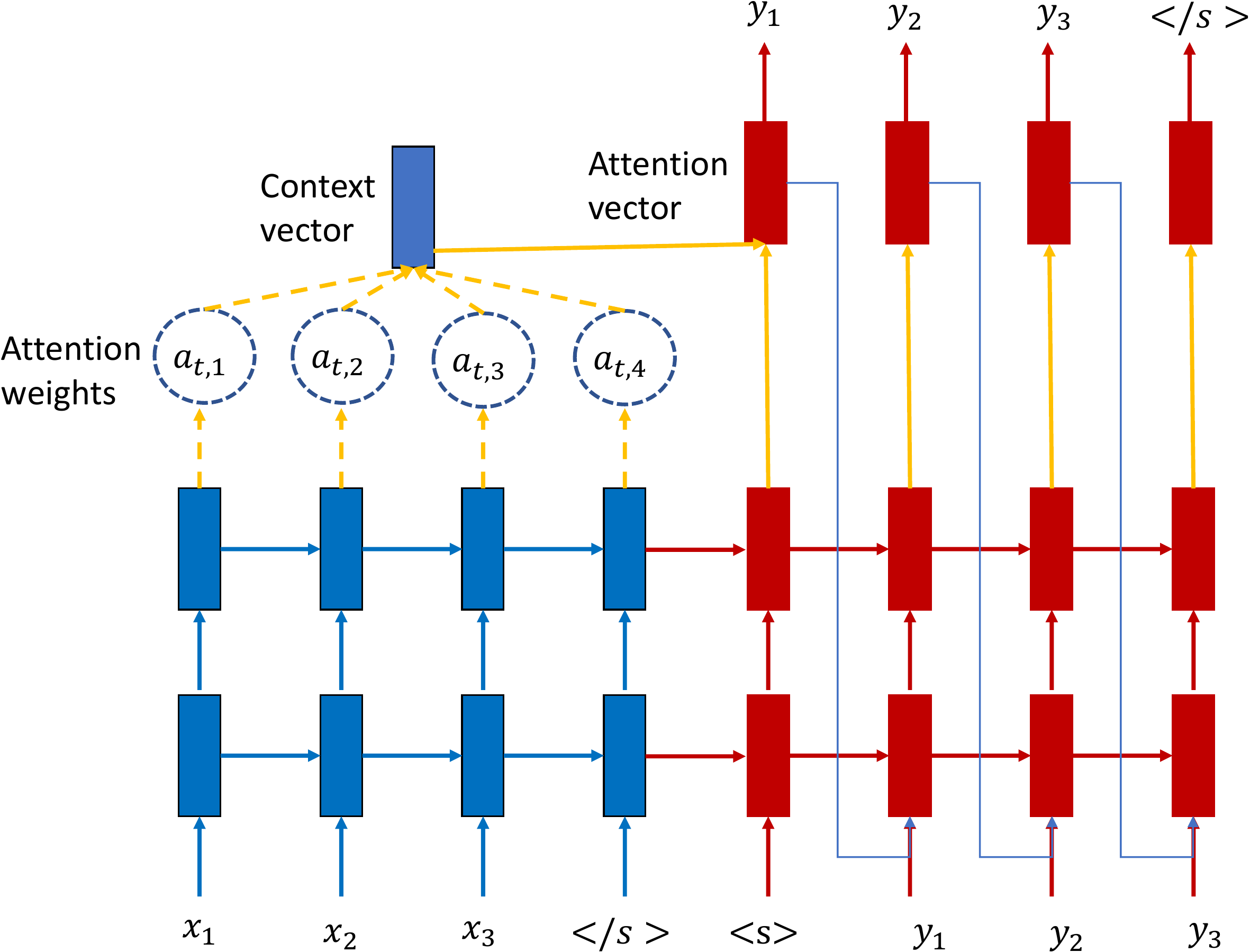}
    \caption{\tnmt{}{\tiny/RNN}}
    \label{fig:tnmt-architecture}
  \end{subfigure}%
  \begin{subfigure}{0.33\linewidth}
    \centering
    \includegraphics[width=0.85\columnwidth]{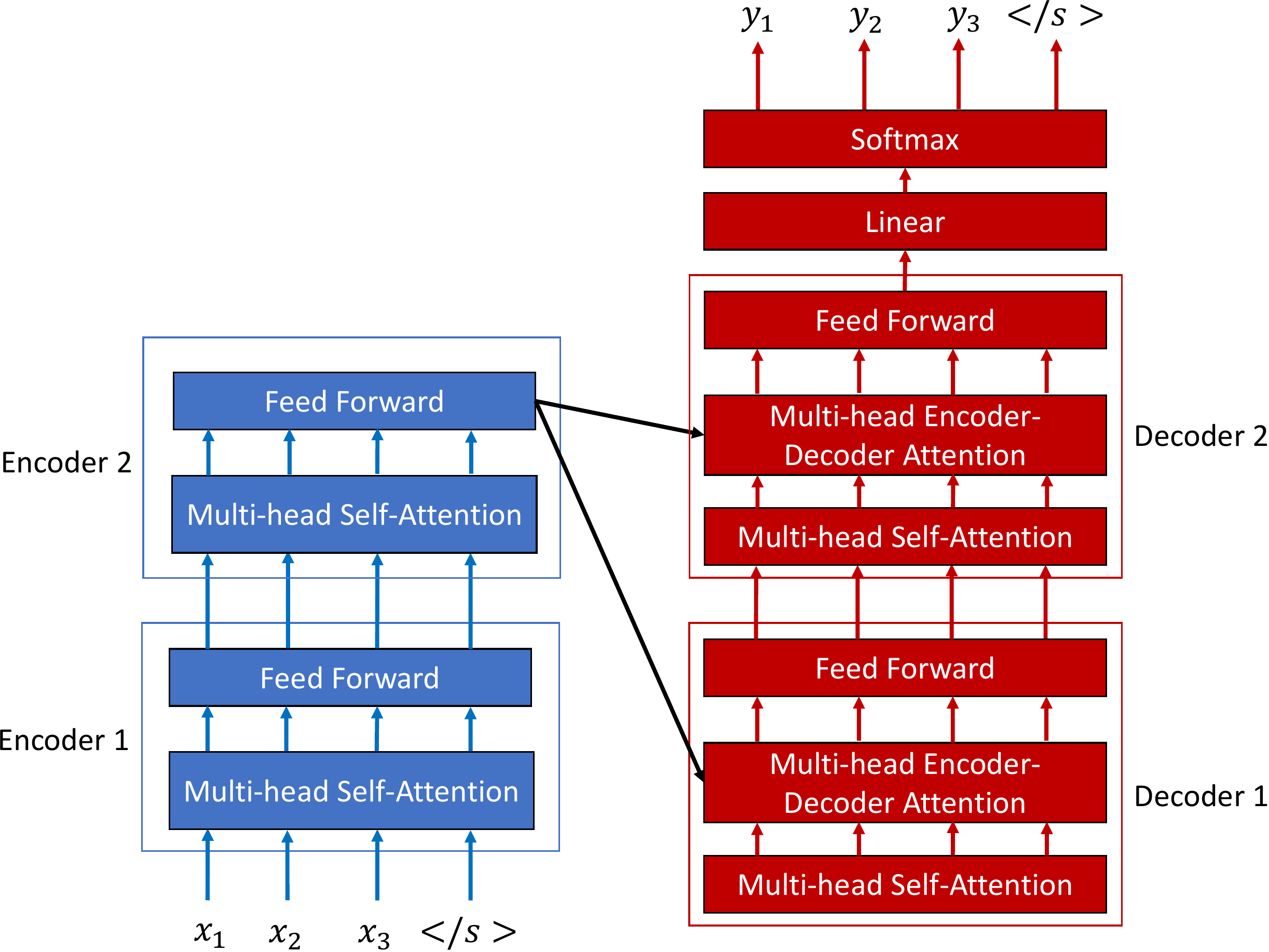}
    \caption{Two-layer Transformer (simplified)}
    \label{fig:simple-transformer-architecture}
  \end{subfigure}
  \caption{Architecture of the three models.}
  \label{fig:arch}\figbotspace
\end{figure*}

\section{Models}
\label{sec:models}
To compare the performance of established and new models, we selected a set of networks consisting of two that use a traditional seq-to-seq architecture with recurrent neural network (RNN) units~\cite{white2018testnmt, watson2020on}, and a more efficient variant of the newer Transformer architecture called \reformer{}~\cite{kitaev2020reformer}.
All of these models utilise the encoder-decoder with attention architecture type where the encoder encodes the source sequence into a vector representation which is then decoded into a target sequence by the decoder. The attention mechanism allows for modelling of out-of-sequence dependencies by attending over the whole source sequence.
This is the typical architecture used for sequence to sequence learning tasks, such as neural machine translation.

\subsection{RNN models}
\label{subsec:rnn-models}
The two RNN type models we use are \atlas{}~\cite{watson2020on} and \tnmt{}~\cite{white2018testnmt}.
For both of these models, the encoder builds the vector representation of the source sequence by traversing the sequence one token at a time converting each token into a vector embedding via an embedding layer which is then provided as input to the encoder RNN unit for that time step.
After the source sequence has been fully processed, the final hidden state of the encoder RNN is used to initialise the hidden state of the decoder RNN.
Then, at each time step, the decoder RNN uses the current hidden state, the previously generated target sequence token, and the attention mechanism, to generate a new target sequence token.
This continues until the end-of-sequence token is generated.

The attention mechanism assists in determining the next token by assigning attention weights to each of the tokens in the source sequence, computing a context vector representing the full attention, and combining them with the hidden state of the decoder to compute the attention vector.
The attention weight $\alpha_{ts}$ for a given target token and source token is computed by performing a normalised comparison between the target hidden state $\mathbf{h}_t$ and the source hidden state $\overline{\mathbf{h}}_s$ using the score function:

\[\alpha_{ts} = \frac{\exp{\textnormal{score(}\mathbf{h}_t, \overline{\mathbf{h}}_s\textnormal{)}}}
{\sum_{s'=1}^{S} \exp{\textnormal{score(}\mathbf{h}_t, \overline{\mathbf{h}}_{s'} \textnormal{)}}}\]

These attention scores are used to compute the context vector $\mathbf{c}_t$ using a weighted sum $\mathbf{c}_t = \mathbf{\sum_{s}} \alpha_{ts} \overline{\mathbf{h}}_s$ and the attention vector is computed by combining the context vector with the current decoder hidden state $\mathbf{a}_t = \textnormal{tanh(}\mathbf{W}_c[\mathbf{c}_t;\mathbf{h}_t]\textnormal{)}$.

The attention vector is then passed to the softmax layer to generate the predicted target token.
After decoding the target token for the current step, the attention vector is passed to the next step in the decoder to ensure that past attention information is carried forward.
This helps to capture contextual and out-of-sequence dependencies by allowing the network to attend to the source tokens in differing amounts as the target sequence is generated.

While the \atlas{} and \tnmt{} networks both utilise this same basic architecture, as shown in \Cref{fig:atlas-architecture} and \Cref{fig:tnmt-architecture}, and both utilise LSTM cells with the tanh activation function, they do differ in several significant ways.
One major difference is that the \atlas{} network includes a copy mechanism~\cite{gu2016incorporating} that replaces UNK token predictions with a token from the source sequence.
In contrast, the \tnmt{} network does not use an UNK replacement mechanism.
Another difference is that the \tnmt{} network uses two unidirectional layers in the encoder and two layers in the decoder, whereas the \atlas{} network uses a single bidirectional layer in the encoder and two layers in the decoder.
This difference between the networks can provide some insight as to the relative effect of the directionality of layers vs the number of layers.
The networks also differ in the way that attention is calculated.
\atlas{} uses Bahdanau's additive technique~\cite{bahdanau2014}: 

\[\textnormal{score(}\mathbf{h}_t, \overline{\mathbf{h}}_s\textnormal{)} = \mathbf{v}_a^\top \textnormal{tanh(}\mathbf{W}_1\mathbf{h}_t + \mathbf{W}_2\mathbf{\overline{h}}_s \textnormal{)}\] 

\tnmt{} uses Luong's multiplicative style~\cite{luong2015}: 

\[\textnormal{score(}\mathbf{h}_t, \overline{\mathbf{h}}_s\textnormal{)} = \mathbf{h}_t^\top \mathbf{W} \overline{\mathbf{h}}_s\]

\subsection{\reformer{} model}
\label{subsec:reformer-model}

The \reformer{} model~\cite{kitaev2020reformer} is a less resource-intensive iteration of the recently popularised Transformer model~\cite{vaswani2017}.
The Transformer model differs from the RNN based models in that it relies solely on attention and simple point-wise fully connected feedforward network layers.
The Transformer still employs encoder-decode attention, however, it also utilises another form of attention called multi-headed self-attention.
The Transformer architecture is comprised of a series of layers stacked on top of each other where each layer contains an encoder and a decoder.
The source sequence is fed through each encoder sequentially and the result is given to each decoder along with the output from the previous decoder (if one exists).
The encoders and decoders are themselves comprised of sub-layers with the encoders containing multi-head self-attention and feedforward sub-layers, while the decoders contain multi-head self-attention, multi-head encoder-decoder attention, and feedforward sub-layers.
The output from the final decoder passes through a single linear layer and into the softmax to compute the output token predictions.
\Cref{fig:simple-transformer-architecture} shows a high-level example of a two-layer Transformer architecture.

The multi-headed attention mechanism works to improve performance by allowing the model to attend to information from multiple different representation subspaces concurrently, enhancing the model's ability to focus on different positions.
This is done by projecting the information from the input vectors $h$ times, where $h$ is the number of heads, performing the attention calculations over each head, and then combining the results from all heads.
All heads are initialised randomly and trained with random dropout so that different heads learn to attend more appropriately over different positions, making the combination of multiple heads more effective than a single attention function.

However, although Transformers achieve state-of-the-art performance they can be very resource-intensive due to the extreme number of parameters and the size of the calculations required for multi-head attention.
Given this limitation, the \reformer{} model was created to reduce the resource requirements of the model while still applying the concepts that make the Transformer effective.
To do this, \reformer{} targets the three main sources of resource consumption in the Transformer, specifically the large self-attention computation, which is $O{L^2}$ for sequences of length $L$, the large numbers of layers, and that the feedforward layers are often much deeper than the attention activations.
The \reformer{} deals with the size of the attention computation by employing Locality Sensitive Hashing (LSH) attention and deals with the large number and depth of layers by using a Reversible Residual Network (RevNet)~\cite{gomez2017deep} with chunking.
However, as the current implementation of the \reformer{}~\cite{google-trax} does not use LSH attention for encoder-decoder sequence to sequence tasks (only decoder-only language models), we omit discussion of LSH here and focus on RevNet and chunking.

\begin{figure}
  \centering
  \includegraphics[width=0.9\columnwidth]{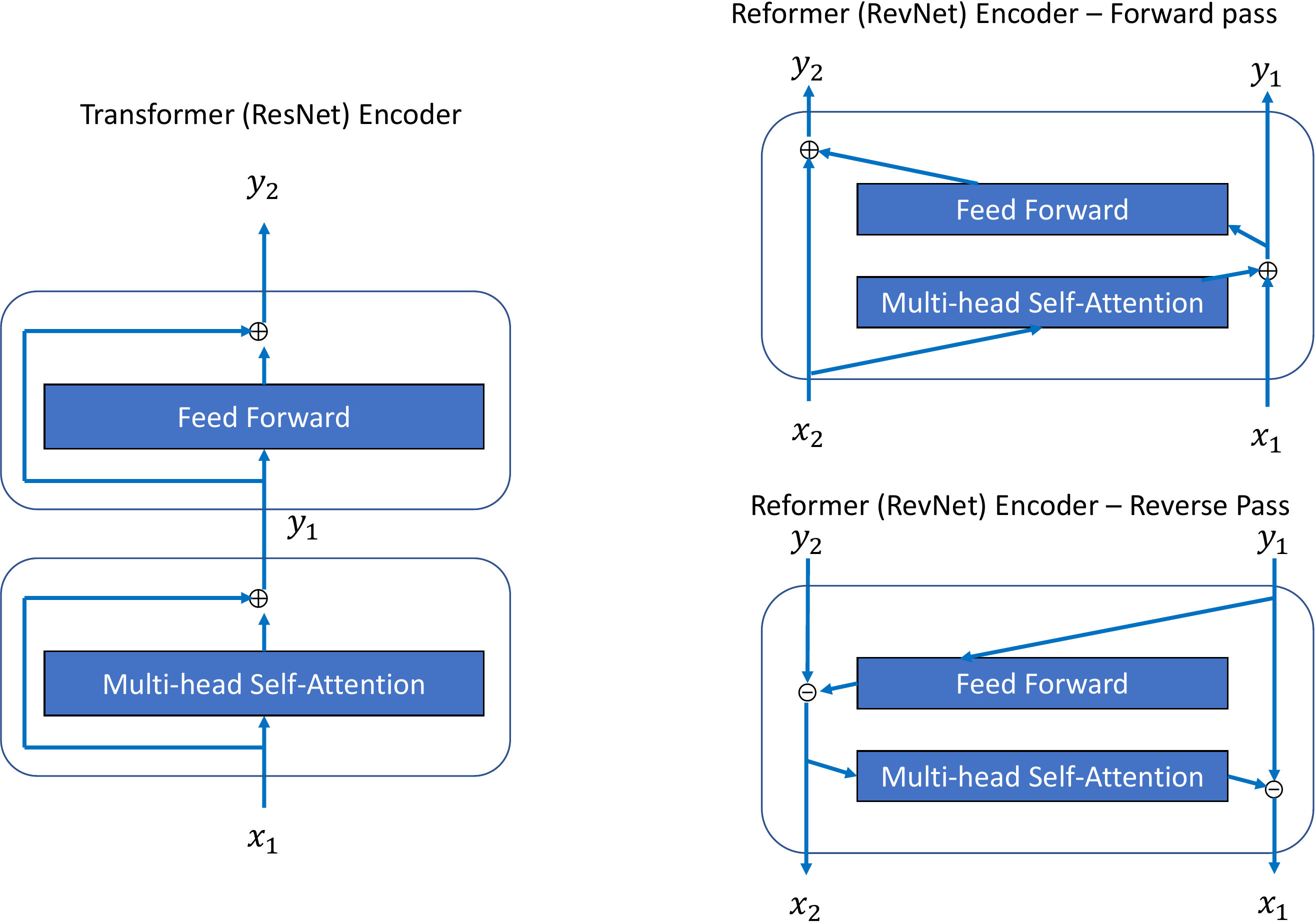}
  \vspace{-1ex}
  \caption{Residual Network (ResNet) units (used in Transformer) vs Reversible Network (RevNet) units (used in \reformer{}).}
  \label{fig:resnet-vs-revnet}\figbotspace
\end{figure}

RevNet improves the memory consumption of the model by replacing the Residual Network (ResNet) units of the standard Transformer with RevNet units.
This reduces memory consumption as ResNet units need to store all of the activations for each layer in memory for the backpropagation calculations, requiring a lot of space when using deep networks.
RevNet units, however, can recover the activations in the $\textrm{n}^\textrm{th}$ layer from the activations in the $\textrm{(n+1)}^\textrm{th}$ layer, meaning only the activations from one layer need to be stored at any time.
\Cref{fig:resnet-vs-revnet} provides an example of this by comparing a Transformer encoder with ResNet units and a \reformer{} encoder with RevNet units.
In traditional transformers, the self-attention and feedforward sub-layers are each contained in their own ResNet units and require that all the activations be stored.
However, in the \reformer{}, the self-attention and feedforward layers are wrapped together in a single RevNet unit where the input (activations from the previous layer) can be recovered from the outputs (activations from the current layer).

The final improvement is using chunking in the feedforward sub-layers.
Improving the efficiency of these layers is important as the dimensionality of the vectors in these layers can reach 4K or higher.
Chunking can be used because the computations are independent across the positions in a sequence, meaning that the computation can be split into $n$ chunks which are executed in series, reducing memory requirements.

By using this combination of efficiency improvements, the \reformer{} can achieve performance on par with that of traditional large Transformers while being much more memory-efficient and faster, especially on large sequences.

\section{Evaluation -- \reassert{}}
\label{sec:evaluation-reassert}
We evaluate the \reassert{} approach using the two RNN-based models from \tnmt~\cite{white2018testnmt} and \atlas~\cite{watson2020on} in addition to the new \reformer{} model~\cite{kitaev2020reformer}.
The projects that we selected to perform the evaluation are Apache OpenNLP~\cite{opennlp}, Deep Learning for Java (DL4J)~\cite{deeplearning4j}, Efficient Java Matrix Library (EJML)~\cite{ejml}, ND4J~\cite{nd4j}, and Stanford CoreNLP~\cite{corenlp}.
These projects are well-tested, widely used, and include two natural language processing libraries (Apache OpenNLP and Stanford CoreNLP), two linear algebra libraries (EJML and ND4J), and one deep learning library (DL4J).
The details of the data sets obtained from these projects are given in \Cref{tab:subjects}.

\begin{table}
  \caption{Subject project details.}
  \centering
  \label{tab:subjects}
  \begin{tabular}{lcS[table-format=5.0]S[table-format=3.0]S[table-format=3.0]}
    \toprule
    Project & Version & \multicolumn{3}{c}{Set Sizes} \\
      & & {Training} & {Validation} & {Test} \\
    \midrule
    Apache OpenNLP & 1.9.1 & 2027 & 66 & 66\\
    DL4J & 1.0.0 & 2122 & 67 & 68\\
    EJML & 0.38 & 26999 & 100 & 100\\
    ND4J & 1.0.0 & 5728 & 49 & 50\\
    Stanford CoreNLP & 3.9.2 & 3930 & 100 & 100\\
    \bottomrule
  \end{tabular}
\end{table}

The evaluation of \reassert{} is split into two research questions which collectively evaluate the usefulness of the generated asserts by performing a lexical accuracy and a dynamic analysis over individual asserts and the applicability of these asserts to the test suites of the projects.

\subsection{Research Question 1 (Assert Accuracy)}
\label{subsec:rq1}
\textit{How many of the generated asserts are exact matches, passing, and compiling?}
In RQ1 we examine the effectiveness of \reassert{} at generating individual asserts when paired with each of the three models.
We perform an analysis on the generated asserts that first establishes which are exact matches (the generated assert exactly matches an assert written by the developers), then which of the remaining asserts compile and which of those then pass when used to replace the developer written asserts in the existing test.

\paragraph{Experimental Setup}
\label{para:rq1-experimental-setup}
To evaluate \reassert{}, we first take each test-to-tested-method pair from the test set, provide the tested method as input to the model, get the output sequence, process the output sequence into syntactically correct assert statements, and compare those statements to those given in the test method.
Where a generated assert exactly matches any assert in the test, we mark it as an exact match (and therefore also as passing and compiling).
We can automatically categorise exact matches in this way as the test suites are fully green (have no failing tests) for all of the projects.
For generated asserts that are not exact matches, we take the test from the pair, remove all existing asserts from the test method, insert the non-matching assert at the end of the test method, attempt to compile and, if compilation is successful, run the test to see if it passes.
We repeat this process for all test-to-tested-method pairs in the test sets of all the projects.

\paragraph{Findings}
\label{para:rq1-findings}
The results, presented in \Cref{tab:rq1-results}, show that, in general, the three models perform similarly.
However, there are some noticeable trends, such as the \tnmt{} model being slightly higher for F1 score in most projects and the \reformer{} model being slightly lower in some projects (precision, recall, and F1 scores are for exact matches only).
Note that in most cases, there are more asserts that pass than asserts that are exact matches and there are more asserts that compile than asserts that pass (i.e., there are some asserts generated that compile, but where the test fails).
A discussion exploring the implications of these results can be found in \Cref{subsec:discussion-assert-accuracy}.
\begin{table}
	\centering
	\caption{RQ1 -- Exact match, passing, and compiling asserts.}
	\label{tab:rq1-results}\tabspace
\begin{tabular}{l@{}S[table-format=3.0]S[table-format=3.0]S[table-format=3.0]S[table-format=3.0]S[table-format=3.0]}
	\toprule
    & \begin{tabular}{@{}c@{}}Apache\\OpenNLP\end{tabular} & \begin{tabular}{@{}c@{}}DL4J\end{tabular} & \begin{tabular}{@{}c@{}}EJML\end{tabular} & \begin{tabular}{@{}c@{}}ND4J\end{tabular} & \begin{tabular}{@{}c@{}}Stanford\\CoreNLP\end{tabular} \\
      \midrule
      \multicolumn{6}{c}{\tnmt{}{\tiny/RNN}}\\
      Gen. Asserts & 173 & 191 & 179 & 118 & 236 \\
      Matches & 45 & 39 & 51 & 10 & 103 \\
      Precision (\%) & 26 & 20 & 28 & 15 & 44 \\
      Recall (\%) & 21 & 13 & 27 & 5 & 30 \\
      F1 & 23 & 16 & 27 & 7 & 35 \\
      Passing & 47 & 40 & 61 & 10 & 108 \\
      Passing (\%) & 27 & 21 & 34 & 15 & 46 \\
      Compiling & 47 & 44 & 69 & 10 & 120 \\
      Compiling (\%) & 27 & 23 & 39 & 15 & 51 \\
      \midrule
      \multicolumn{6}{c}{\atlas{}{\tiny/RNN}}\\
      Gen. Asserts & 150 & 111 & 184 & 96 & 197 \\
      Matches & 47 & 21 & 45 & 9 & 85 \\
      Precision (\%) & 31 & 19 & 24 & 9 & 43 \\
      Recall (\%) & 22 & 7 & 24 & 4 & 25 \\
      F1 & 26 & 10 & 24 & 6 & 31 \\
      Passing & 47 & 21 & 53 & 14 & 87 \\
      Passing (\%) & 31 & 19 & 29 & 22 & 44 \\
      Compiling & 47 & 28 & 63 & 14 & 100 \\
      Compiling (\%) & 31 & 25 & 34 & 22 & 51 \\
      \midrule
      \multicolumn{6}{c}{\reformer{}}\\
      Gen. Asserts & 192 & 217 & 212 & 132 & 271 \\
      Matches & 30 & 17 & 38 & 8 & 88 \\
      Precision (\%) & 16 & 8 & 18 & 6 & 32 \\
      Recall (\%) & 15 & 6 & 20 & 4 & 25 \\
      F1 & 15 & 7 & 19 & 5 & 29 \\
      Passing & 30 & 17 & 49 & 9 & 98 \\
      Passing (\%) & 16 & 8 & 23 & 7 & 36 \\
      Compiling & 31 & 26 & 61 & 9 & 110 \\
      Compiling (\%) & 16 & 12 & 29 & 7 & 41 \\
	\bottomrule
	\end{tabular}
\end{table}

\subsection{Research Question 2 (Test Applicability)}
\label{subsec:rq2}
\textit{What percentage of tests contain at least one assert from the categories?}
In RQ2 we perform an analysis that uses the generated asserts from RQ1 where, for each assert category (exact match, passing, compiling), we determine the percentage of tests that have at least one generated assert from that category.
This is to give evidence as to how useful the generated asserts are across a whole test suite.

\paragraph{Experimental Setup}
\label{para:rq2-experimental-setup}
To answer this research question, we use the asserts generated for RQ1 and, for each category, count the percentage of tests in each project that contains at least one assert from that category. 

\paragraph{Findings}
\label{para:rq2-findings}
The results, presented in \Cref{tab:rq2-results}, show that in the best case, using the \tnmt{} model, nearly half of the tests in a project receive a generated assert that at least compiles (Stanford CoreNLP).
On average, a third of tests receive at least one generated assert that compiles and 28\% receive at least one exact match assert.
When comparing models the performance is similar but, like RQ1, the \reformer{} model is slightly lower in some projects than the two RNN models \tnmt{} and \atlas{}.
\begin{table}
	\centering
	\caption{RQ2 -- Percentage of tests with at least one generated assert that is an exact match, passing, or compiling.}
	\label{tab:rq2-results}\tabspace
	\begin{tabular}{l@{}S[table-format=3.0]S[table-format=3.0]S[table-format=3.0]S[table-format=3.0]S[table-format=3.0]}
	\toprule
    & \begin{tabular}{@{}c@{}}Apache\\OpenNLP\end{tabular} & \begin{tabular}{@{}c@{}}DL4J\end{tabular} & \begin{tabular}{@{}c@{}}EJML\end{tabular} & \begin{tabular}{@{}c@{}}ND4J\end{tabular} & \begin{tabular}{@{}c@{}}Stanford\\CoreNLP\end{tabular} \\
      \midrule
      \multicolumn{6}{c}{\tnmt{}{\tiny/RNN}}\\
      Matched & 26\% & 27\% & 29\% & 20\% & 40\% \\
      Passing & 27\% & 28\% & 32\% & 20\% & 44\% \\
      Compiling & 27\% & 31\% & 37\% & 20\% & 49\% \\
      \midrule
      \multicolumn{6}{c}{\atlas{}{\tiny/RNN}}\\
      Matched & 26\% & 12\% & 32\% & 18\% & 40\% \\
      Passing & 26\% & 12\% & 37\% & 22\% & 40\% \\
      Compiling & 26\% & 18\% & 42\% & 22\% & 44\% \\
      \midrule
      \multicolumn{6}{c}{\reformer{}}\\
      Matched & 17\% & 13\% & 28\% & 16\% & 41\% \\
      Passing & 17\% & 13\% & 33\% & 18\% & 45\% \\
      Compiling & 18\% & 21\% & 38\% & 18\% & 48\% \\
	\bottomrule
	\end{tabular}
\end{table}

\section{Evaluation -- \atlas{}}
\label{sec:evaluation-atlas}
We present our evaluation of the three models using \atlas{} to determine if we can improve over previous results using the new \reformer{} model or the \tnmt{} model.
\atlas{} significantly differs from \reassert{} in two aspects: (a) \atlas{} uses the tested method and the test method for training and querying and (b) \atlas{} uses a very simple and imprecise test-to-code traceability technique. However, it has been applied in a multi-project setting in which the corpus is created from a large number of projects.
To construct the data set, Watson et al.\ mined 9,275 projects from GitHub and used the Spoon library~\cite{pawlak2016spoon} to extract the test methods by looking for the \textit{@Test} annotation.
However, any test that contained more than one assert statement or was longer than 1000 tokens was discarded, leaving 188,154 tests in total.

The traceability technique used by Watson et al.~\cite{watson2020on} in \atlas{} is a simplified version of Last Call Before Assert (LCBA)~\cite{rompaey2009establishing}. Instead of using a static or dynamic call graph, \atlas{} simply extracts the name of the last called method before the assert and then searches the package for methods of the same name.
If no match can be found, \atlas{} extends the search to the whole project.
While having the benefit of being able to be used on a large and diverse corpus, this method for establishing test-to-code traceability links can result in a lot of noise in the data.
The noise can be especially bad if there are multiple classes which define methods with the same names or if there are a lot of overloaded methods.
After establishing the links, \atlas{} processes them into input-output examples by extracting the asserts from the tests to use as the outputs with their respective tested methods as the inputs.
Further filtering is then performed on the resulting data set to remove duplicate examples and any example where the assert contains a token that does not appear in the vocabulary.
The data set provided by Watson et al.\ is already filtered, so our evaluation uses the data set directly without mining or extraction.

The evaluation of the three models using \atlas{} is split into three research questions which collectively evaluate the usefulness of the generated asserts by looking at the accuracy (RQ3 and RQ4) and uniqueness (RQ5) of the generated asserts.

\subsection{Research Question 3 (Assert Accuracy)}
\label{subsec:rq3}
\textit{How many of the generated asserts are exact matches for developer written asserts?}
For RQ3 we examine the effectiveness of our three models at generating exact match asserts, similar to RQ1, but in the \atlas{} setting.
The evaluation is limited to exact matches as performing a dynamic analysis to discover passing or compiling non-matched asserts is not possible with Watson et al.'s data set.
We do not use beam search when applying the \atlas{} model as it results in multiple tokens being predicted for the same position in the output sequence.
Therefore, when it is utilised in the same way as Watson et al.\ and all of the predicted tokens are used to build a list of possible outputs, the output of the model is a set of candidate assert recommendations rather than a single assert.

\paragraph{Experimental Setup}
\label{para:rq3-experimental-setup}
To answer RQ3, we use the model to generate an assert for each test-to-tested-method pair in the test set and compare the generated assert to the assert from the test as present in the data set.
Where the generated assert and the test assert match, we count it as an exact match and use the number of exact matches divided by the total number of generated asserts to calculate the precision.

\paragraph{Findings}
\label{para:rq3-findings}
The results, as shown in \Cref{tab:rq3-results}, reveal that \tnmt{} is the worst-performing model with only 7\% precision. While \atlas{} fairs much better than \tnmt{} with 17\% precision, \reformer{} is the best by a wide margin at 28\% precision.
Note that our results of 3323 exact matches for our reimplementation of \atlas{} is identical to the results reported by Watson et al.~\cite{watson2020on}, giving us confidence that our reimplementation is faithful to the original \atlas{}.
Discussion regarding these results can be found in \Cref{subsec:discussion-assert-accuracy}. 
\begin{table}
	\centering
	\caption{RQ3 -- Exact match asserts.}
	\label{tab:rq3-results}\tabspace
	\begin{tabular}{lrrr}
	\toprule
    & \tnmt{}{\tiny/RNN} & \atlas{}{\tiny /RNN} & \reformer{} \\
      \midrule
      Generated Asserts & 18\,817~~~ & 18\,817~~ & 18\,817~ \\
      Matched Asserts & 1355~~~ & 3323~~ & \textbf{5262}~ \\
      Accuracy & 7\%~~~ & 18\%~~ & \textbf{28\%}~ \\
	\bottomrule
	\end{tabular}
\end{table}

\subsection{Research Question 4 (Edit Distance Evaluation)}
\label{subsec:rq4}
\textit{How far from exact matches are non-matched asserts?}
RQ4 investigates how much transformation, measured in absolute and relative token-based edit distance, is required to turn non-exact match asserts into exact matches.
These measures give evidence as to how useful non-matched asserts are to developers as, intuitively, the easier it is to turn a non-exact match assert into an exact match, the more useful that assert would be for developers.
We use the relative edit distance as we want to take the length of the asserts into account to avoid favouring models that are more likely to produce short asserts.
Discussions relating to the length of generated asserts can be found in \Cref{subsec:discussion-assert-uniqueness}.
We also report the count of asserts that are less than two token changes away from being a matched assert. 
This group, therefore, includes asserts that are either exact matches or only one token change away from an exact match.
Given the ease of changing a single token, we consider these non-matched asserts to be in the group of asserts which should be of most use to developers.

\paragraph{Experimental Setup}
\label{para:rq4-experimental-setup}
This evaluation is performed using the asserts generated for RQ5.
First, we find the edit distance by computing the Levenshtein distance between the generated assert and each developer written assert, using tokens instead of characters as the atomic unit, and take the smallest distance.
The distance is then used to compute the relative edit distance by dividing it by the number of tokens in the assert with the most tokens out of the generated assert and the developer written assert.

\paragraph{Findings}
\label{para:rq4-findings}
The results, as shown in \Cref{tab:rq5-results} reveal that the \atlas{} and \reformer{} models perform essentially equivalently to each other in edit distance, with the \tnmt{} model trailing behind them.
However, when looking at asserts that are less than 2 token changes away from an exact match, the \reformer{} model has a clear advantage.
\begin{table}
	\centering
	\caption{RQ4 -- Exact match edit distance evaluation.}
	\label{tab:rq4-results}\tabspace
	\begin{tabular}{l@{}rrr}
	\toprule
    & \tnmt{}{\tiny/RNN} & \atlas{\tiny/RNN} & \reformer{} \\
      \midrule
      Median Edit Dist. & 5~~~ & \textbf{2}~~~ & \textbf{2}~~ \\
      Mean Edit Dist. & 5.07~~~ & \textbf{3.85}~~~ & 4.00~~ \\
      Median Rel. Edit Dist. & 0.28~~~ & \textbf{0.15}~~~ & \textbf{0.15}~~ \\
      Mean Rel. Edit Dist. & 0.26~~~ & \textbf{0.18}~~~ & 0.19~~ \\
      Dist. $< 2$ Count & 3375~~~ & 6984~~~ & \textbf{8180}~~ \\
      Dist. $< 2$ (\%) & 18\%~~~ & 37\%~~~ & \textbf{43\%}~~ \\
	\bottomrule
	\end{tabular}
\end{table}

\subsection{Research Question 5 (Uniqueness Evaluation)}
\label{subsec:rq5}
\textit{What is the uniqueness of generated asserts?}
RQ5 investigates how unique the generated asserts are,
which is important as the more unique an assert is, the more useful it is likely to be.
This belief is driven by the fact that, in general, asserts that are more unique are more likely to encode specific information about the task.
For example, an assert statement that simply checks the equality of two generically named variables contains less specific information than an assert statement which contains method calls.
We use the asserts generated by each of the three models only with the Watson et al.\ data set because this data set is taken from a large number of projects and, therefore, demonstrating the ability to generate a diverse and unique range of asserts is important.

To evaluate uniqueness, we first look at the absolute number of unique asserts the models produce and what percentage of generated asserts were unique at generation time for all generated asserts and all matched asserts.
This measures how frequently the models are generating unique asserts.
However, we do not only want to look at unique asserts but also the distribution of non-unique asserts.
We perform this analysis with a view that a more even distribution, in general, indicates a greater diversity of asserts and, therefore, greater useful informational content.
This assumption is discussed in more detail in \Cref{subsec:discussion-assert-uniqueness}.
To assess the distribution of non-unique asserts, we compute the absolute number and percentage of matched asserts that are among the top five and top ten most common asserts, essentially showing us how common the most common asserts are.
To demonstrate a good ability to generate asserts with a high degree of uniqueness, we are looking for a model to maximise the unique assert percentages while minimising the most common assert percentages.

\paragraph{Experimental Setup}
\label{para:rq5-experimental-setup}
To conduct RQ5, for each model, we first take the list of assert statements generated by the model and group identical asserts together.
The sizes of these groups give us the count of how many times each assert appears.
We take the number of groups as our count of distinct asserts and calculate this as a percentage of all the generated asserts.
This is the percentage of asserts that were unique at the time of generation.
The groups are then ordered by their cardinalities and we take the sum of the cardinalities of the top five and the top ten largest groups and use these to calculate the percentage of generated asserts that are members of these groups.

\paragraph{Findings}
\label{para:rq5-findings}
The results, as shown in \Cref{tab:rq5-results} reveal that \reformer{} is the best model for uniqueness as it has the highest percentages of unique asserts and the lowest percentages of asserts that are among the top 5 and top 10 most common asserts.
These results show \reformer{} is better for uniqueness than the next best model, \atlas{}, by a sizeable margin in all measures.
The \tnmt{} model performs poorly as it rarely generates unique asserts.
Discussion regarding these results can be found in \Cref{subsec:discussion-assert-uniqueness}. 
\begin{table}
	\centering
	\caption{RQ5 -- Assert uniqueness analysis results.}
	\label{tab:rq5-results}\tabspace
	\begin{tabular}{l@{}rrr}
	\toprule
    & \tnmt{}{\tiny/RNN} & \atlas{}{\tiny/RNN} & \reformer{} \\
      \midrule
      Unique Asserts & 470~~~ & 11\,496~~ & \textbf{13\,331}~ \\
      Unique Asserts (\%) & 2\%~~~ & 62\%~~ & \textbf{71\%}~ \\
      Unique matched & 97~~~ & 948~~ & \textbf{2647}~ \\
      Unique matched (\%) & 7\%~~~ & 29\%~~ & \textbf{50\%}~ \\
      Top 5 matched (\%) & 59\%~~~ & 25\%~~ & \textbf{16\%}~ \\
      Top 10 matched (\%) & 71\%~~~ & 37\%~~ & \textbf{20\%}~ \\
	\bottomrule
	\end{tabular}
\end{table}

\section{Discussion}
\label{sec:discussion}
We discuss the findings of the research questions and other subjects relating to our methodology and outcomes.
The topics of assert accuracy (RQ1 -- RQ4) and assert uniqueness (RQ5) are of particular interest as they constitute the primary ways in which we assess the usefulness of the generated asserts.
In addition, there are important takeaway messages regarding the practicalities of applying this general approach to code generation tasks, both in research and in industrial practice.

\subsection{Assert Accuracy}
\label{subsec:discussion-assert-accuracy}
Assessing the accuracy of the generated asserts by comparing to a ground truth test set is the primary method for evaluating assert generation techniques as it shows us how similar the generated asserts are to developer written asserts. 
Given the assumption that developers write useful asserts, this gives direct evidence for the usefulness of the generated asserts.

For RQ1 we use the precision and recall as our measure for accuracy which shows that the accuracy achieved by \reassert{} is heavily dependant on the project it is applied to.
In the best case of our experiments, using the \tnmt{} model with Stanford CoreNLP, the accuracy is greater than what is achieved in the best case with \atlas{}, using the \reformer{} model.
However, when using the ND4J project, the accuracy is lower.
Despite this, when taking the RQ2 results into consideration, we see that even for the worst-performing project, ND4J, we still have 20\% of tests receiving at least one exact match assert.

When using the accuracy to compare models within the \atlas{} approach, in RQ3 we see that the \reformer{} model with 5262 matched asserts is 58\% more accurate than the \atlas{} model with 3323 matched asserts, the next best performing model, while the \tnmt{} model is far behind with only 1355 matched asserts.
The poor performance of \tnmt{} is due to its lack of an UNK replacement mechanism which results in an UNK token appearing in 80\% of the asserts it generates.
As any assert which contains an UNK token cannot be matched, the accuracy of the model is very poor.
This is one of the primary ways in which the \atlas{} model differs from the \tnmt{} model in that it implements a copy mechanism that replaces UNK token predictions with a token from the source sequence, the effects of which are seen in these results.
RQ1 paints a different picture in terms of the comparisons between models when using the \reassert{} approach.
This that shows that all the models are close to each other in general but where there are larger differences, the ordering from RQ1 is typically reversed, with \tnmt{} coming in first and \reformer{} coming in last.
The reasons for this are two-fold.
Firstly, it seems that the accuracy of the models is ultimately being bottlenecked by the quantity and diversity of training data.
Given that the amount of training data that can be extracted from a single project is limited, there are a proportion of asserts appearing in the test set that bare no resemblance to any assert in the training set and, therefore, will never be able to be replicated by any model.
Given this, it seems that \tnmt{} may be the best model at learning and recreating a restricted set of asserts that appear frequently, while \reformer{} is better at generalising when given a more diverse data set.
This would explain the differences between these models when comparing the RQ1 and RQ3 results.
The subject of the effect of data sets on the performance of the models is discussed further in \Cref{subsec:data-set-size-and-diversity}.

The takeaway message from these RQs is that the best performing model is dependent on the usage scenario.
If generating asserts for a project that has a data set that is conducive to sequence to sequence learning, \reassert{} with a \tnmt{} model is the best performing with up to 44\% precision and the ability to generate at least one matching assert for up to 40\% of tests with the projects we used in the evaluation.
Otherwise, when using the \atlas{} approach, the \reformer{} model may be the best choice.

\subsection{Assert Uniqueness}
\label{subsec:discussion-assert-uniqueness}
\begin{table}
	\centering
	\caption{Top 5 most common matched asserts across all models.}
	\label{tab:common-asserts}\tabspace
	\begin{tabular}{lr}
	\toprule
    Assert & Count \\
      \midrule
      assertEquals(expected, actual) & 1288 \\
      assertEquals(expResult, result) & 419 \\
      assertEquals(expected, result) & 337 \\
      assertTrue(true) & 253 \\
      assertNotNull(result) & 181 \\
	\bottomrule
	\end{tabular}
\end{table}
\begin{table}
	\centering
	\caption{Top 5 most common matched asserts containing a method call.}
	\label{tab:common-method-call-asserts}\tabspace
	\begin{tabular}{lr}
	\toprule
    Assert & Count \\
      \midrule
      assertEquals(0, result.size()) & 172 \\
      assertTrue(getNoErrorMsg(), result) & 59 \\
      assertEquals(0, meldingen.size()) & 57 \\
      assertEquals(200, response.getStatusCode()) & 38 \\
      assertEquals("test", echo.echo("test")) & 26 \\
	\bottomrule
	\end{tabular}
\end{table}

In RQ5, we performed a uniqueness evaluation on the generated asserts to provide more evidence for how useful the asserts are in practice.
This was done as uniqueness is an indicator of specificity and the more specific information an assert contains, the more useful that assert is likely to be to in practice.
Therefore, we use uniqueness as a partial proxy for evaluating usefulness.
The intuition behind this evaluation is clear when inspecting the least unique (most commonly generated) asserts, as shown in \Cref{tab:common-asserts}.
This demonstrates how the most common asserts are extremely generic and provide almost no specific information to the developers as they simply compare values of generically named variables.
In the extreme case, as exemplified by the fourth most common assert, \textit{assertTrue(true)}, the assert is of no use at all and has been learnt from developers writing a placeholder assert into their tests (which is considered bad practice).
As a comparison, if we look at the top five most common generated asserts that contain a method call, as shown in \Cref{tab:common-method-call-asserts} we see that, while still somewhat generic, these asserts contain more specific information for how to test the tested method.
This comparison highlights how uniqueness relates to specificity, which in turn relates to practical usefulness.
We, therefore, favour models which generate the greatest diversity of asserts.

Given that \reformer{} produces more unique asserts and has a lower percentage of its asserts belonging to the top 5 and top 10 most common asserts as compared to \atlas{} in the evaluation for RQ5, \reformer{} is the most desirable model in this regard.
\tnmt{} performs poorly in this evaluation for the same reason as its poor performance in accuracy, namely that the lack of an UNK replacement mechanism limits the range of matched asserts that it can produce.
The takeaway message is that the use of a state of the art model like \reformer{} can improve the usefulness of the generated asserts due to the higher uniqueness of the asserts.

\subsection{Data Set Size, Diversity, and Quality}
\label{subsec:data-set-size-and-diversity}
As discussed in \Cref{subsec:discussion-assert-accuracy}, we see a surprising result when we compare the accuracy between models when using \reassert{} versus when using \atlas{}, in that the models perform much more similarly with \reassert{}.
This indicates that some of the projects selected for the \reassert{} evaluation produce data sets that do not allow all the models to generalise maximally, most likely due to insufficient size, low diversity, or too much noise.
This is an important takeaway for those wishing to use these code generation techniques in the future as these properties are determined not just by the size of the projects from which the data is taken but also by the traceability technique used to establish the test-to-tested-method links, the filtering that is applied to the data sets, and the way the code is written.
This diversity of concerns is evident when investigating the relationship between data set sizes and performance.
In this regard, it's important to note that the project with the largest data set (EJML) is only the second-best performing project in terms of F1 score in RQ1 with the \reformer{} model (and third-best with the RNN models), while the project that performs best for F1 score with all models (Stanford CoreNLP) has only the third largest data set.
This shows that the size, diversity, and quality of the data set has a large impact and, for projects of this size, ultimately limits the ability of the models to generalise.
The takeaway message, therefore, is that it is crucial to select a corpus of software that is large and diverse and that appropriate techniques which balance data set size and accuracy must be selected.

\section{Threats to Validity}
\label{sec:threats-to-validity}
The threats to validity are related to the data used for training and evaluating the models, both in terms of subject selection and the method of data set collection.
An external threat to validity is the representativeness of the subjects chosen for the \reassert{} evaluation, as we have no strong evidence that the subjects are representative of the general population of software.
However, the subjects cover a range of project types, are widely used in research and industry, and are large enough to demonstrate applicability to complex software. 
The second threat comes from the method by which the data was collected as the traceability techniques used to build the test-to-tested-method links do not have complete precision and there is, therefore, some noise in the data.
However, as discussed in~\Cref{subsec:approach-test-to-code-traceability-establishment}, we believe that having some noise in the data does not necessarily significantly hamper the training.
For the individual project data sets, we ensure that the validation and test sets contain minimal noise by using a very high precision traceability technique for constructing those sets.
However, when using multi-project data sets, such as in the \atlas{} evaluation, the results may vary if using a data set with a significantly different amount of noise in the data sets.

\section{Related Work}
\label{sec:related-work}
Prior to the application of the machine learning techniques that are the subject of this paper, assert generation was done primarily by test suite generation tools.
These tools can be split into several categories depending on the general approach used for the generation of their tests.
Randoop~\cite{pacheco2007randoop}, Nighthawk~\cite{andrews2007nighthawk}, JCrasher~\cite{csallner2004jcrasher}, and CarFast~\cite{park2012carfast} are the primary examples of tools that use approaches based on random generation while
EvoSuite~\cite{fraser2013whole} and eToc~\cite{tonella2004evolutionary} are examples of meta-heuristic search-based tools and Symbolic Pathfinder~\cite{pasareanu2010symbolic} and jCUTE~\cite{sen2006jcute} are examples of tools that use dynamic symbolic execution.
However, despite the diversity of approaches to test generation employed by these tools, they all focus primarily on things other than the generation of meaningful asserts.
The usual goal for these tools is achieving coverage or exposing faults in other ways, such as generating exceptions and crashes. 
Therefore, even the most well developed and studied examples of these tools which do have some form of assert generation, such as EvoSuite and Randoop, the asserts they generate are often trivial or not meaningful, contributing to the relatively high rate of missed faults in real-world projects~\cite{shamshiri2015do}.

\section{Conclusion}
\label{sec:conclusion}
We have presented \reassert{}, a project-based deep learning approach for the generation of JUnit test asserts.
We also utilise the state-of-the-art \reformer{} model and two RNN-based models from previous work to evaluate \reassert{} and provide an extended evaluation of \atlas{}, allowing us to compare models and approaches for assert generation.
\reassert{} improves over previous work by generating asserts that are, in general, more accurate and does not require that a test be written before being able to generate asserts, in addition to being able to generate multiple asserts for a single function.
Also, the \reformer{} model is shown to improve the results achievable by the \atlas{} approach~\cite{watson2020on} generating asserts that are more accurate and more unique.
However, when the \reformer{} model is used with \reassert{}, the difference in effectiveness between the models is greatly lessened.
This indicates that some of the projects selected for the \reassert{} evaluation produce data sets that do not allow all the models to generalise maximally, most likely due to insufficient size, low diversity, or too much noise.
Therefore, researchers and practitioners must be aware of this limitation and select code corpora and traceability techniques that provide suitably large, diverse, and clean data sets.

\bibliographystyle{IEEEtran}
\bibliography{reassert-arxiv}

\end{document}